\documentclass{ws-ijmpb}

\begin{document}
\title{Transport in a Clean Graphene Sheet at Finite Temperature
and Frequency}

\author{N. M. R. Peres$^1$ and T. Stauber$^{1,2}$}
\address{$^1$Center of Physics and Department of
Physics, University of Minho, P-4710-057, Braga, Portugal\\
$^2$Instituto de Ciencia de Materiales de
Madrid. CSIC. Cantoblanco. E-28049 Madrid, Spain}
\maketitle
\begin{abstract}
We calculate the conductivity of a clean graphene sheet at finite temperatures starting from the tight-binding model. We obtain a finite value for the dc-conductivity at zero temperature. For finite temperature, the spontaneous electron-hole creation, responsible for the finite conductivity at zero temperature, is washed out and the dc-conductivity yields zero. Our results are in agreement with calculations based on the field-theoretical model for graphene.
\end{abstract}
\section{Introduction}
The recent achievements of isolating and locating a single layer of graphene and performing transport measurements have tremendously stimulated the research on carbon-based electronics.\cite{Exp1} By applying an external gate voltage, the system can be switched from $n$-type to $p$-type carriers, thus controlling the electronic properties and giving rise to carbon-based devices. At the crossover from $n$- to $p$-type carriers, i.e., at the charge neutrality point where no carriers are present, experiments nevertheless report a minimal finite, ``universal'' conductivity.\cite{Exp2}

A first understanding of this result is already obtained from the band structure of the system. Within a two-band tight-binding approach\cite{McClure} and in the long-wavelength limit, single-layer graphene is described by two-dimensional Dirac Fermions (see e.g. Ref. \cite{Peres06}). At zero doping, the density-of-states (DOS) at the Fermi energy is zero and the system is a zero-band gap semiconductor or semi-metal. Graphene thus lies at the borderline of a semiconductor and a metal which gives rise to a number of new phenomena. E.g., in contrast to the zero (infinite) conductivity for a clean semiconductor (metal), the conductivity of a clean graphene sheet at zero temperature is finite and of order $e^2/h$.

Already Ludwig et. al.\cite{Ludwig94} pointed out that there are two
possible prefactors to the conductance quantum $e^2/h$. Calculating,
e.g., the dc-conductivity via the Kubo-formula using the
polarizability for undoped graphene of Ref. \cite{Gon94} yields
$\sigma=\frac{\pi}{2}\frac{e^2}{h}$, where the dielectric constant
$\epsilon$ is usually omitted ($\epsilon=1$).\cite{FootnoteDeg}
Introducing a finite scattering amplitude $\Gamma$ due to impurities
and performing the limit $\Gamma\to0$ in the end leads to
$\sigma=\frac{4}{\pi}\frac{e^2}{h}$.\cite{Gorbar02}

Besides using the Kubo formula, there were also calculations based on
the Landauer formula for a rectangular system with aspect ratio
$W/L\gg1$.\cite{Katsnelson06,Beenakker06} Independent from the explicit
boundary conditions, the conductivity was given by
$\sigma=\frac{4}{\pi}\frac{e^2}{h}$, corresponding to the physical
limit procedure (first $\omega\to0$ and then $\Gamma\to0$) in the case
of the infinite system without contacts. In Ref. \cite{Ryu06},
Ryu et. al. thoroughly discuss the crossover between the two values
depending on the different dc-limits. But taking the dc-limit after the integration over energies is performed, one obtains a third finite value for the minimal conductivity, i.e., $\sigma=\pi\frac{e^2}{h}$ which is closest to the experimentally measured value given by $\sigma\approx4\frac{e^2}{h}$.\cite{Ziegler06} 

The above works are based on a ballistic transport model which predicts a universal conductivity. But Mishchenko pointed out that electron-electron interaction may lead to a non-universal correction.\cite{Mish06} Also diffusive transport models can account for a minimal, but non-universal conductivity via midgap states,\cite{Sta07} Coulomb scatterers\cite{Tan07} or electron-hole puddles.\cite{Cheianov07}

This paper shall focus on the transport properties of clean graphene
sheets at finite temperature and frequency. For an overview on ac
conductivity in graphene see Ref. \cite{Sharapov07}. The effect of temperature was also discussed in Ref. \cite{Ryu06} which can
have two contrary effects as outlined below.

On the one hand, one can see temperature as some
effective finite chemical potential which would lead to an enhancement
of the conductivity.\cite{FootnoteChemPot} This is suggested by the
results by Vafek on the plasmon dispersion for graphene at zero doping
and at finite temperature \cite{Vafek06}. He finds that the collective
plasmon excitations are only weakly damped even though they lie in the
region where electron-hole excitations are allowed. In the case of
finite doping and zero temperature, electron-hole excitations are
forbidden due to the Pauli principle, thus leading to undamped plasmon
excitations within the RPA approximation.\cite{Hwang06,Wunsch06}

On the other hand, the finite conductivity at $T=0$ can be related to the spontaneous electron-hole creation at zero energy.\cite{Katsnelson06} These quantum fluctuations can be washed out at finite temperature, thus rendering the mechanism for finite conductivity for clean graphene to zero. It is this scenario which prevails and the dc-conductivity for graphene is indeed zero if the energy scale set by the carrier density is smaller than the temperature.

All calculations on the conductivity of clean graphene cited above are based on the long-wavelength approximation, i.e.,  start from relativistic Dirac Fermions. Here, we want to discuss the conductivity of a clean graphene sheet starting from the tight-binding model. We will further include finite temperature since the order of limits $\omega\to0$ and then $T\to0$ corresponds to the physical realization of an experiment. 

\section {Tight-binding model with vector potential}
The simplest Hamiltonian describing non-interactive electrons on
graphene in the presence of a time dependent vector potential $\bm A(t)$ reads 
\begin{eqnarray}
H  =  -\sum_{\mathbf{R},\sigma,\bm\delta }[t_{\bm R,\bm R+\bm\delta}
a_{\sigma}^{\dagger}
(\mathbf{R})b_{\sigma}(\mathbf{R}+\bm\delta)+\textrm{H.c.}]\,,
\label{eq:Hlayer}
\end{eqnarray}
where $\bm R$ runs over all unit cells and $\bm \delta$ runs over all three nearest neighbors with 
\begin{eqnarray}
\bm \delta_1 &=& \frac a 2 (-1,\sqrt 3,0),\\
\bm \delta_2 &=& \frac a 2 (-1,-\sqrt 3,0), \\
\bm \delta_3 &=& a (1,0,0).
\label{nnv}
\end{eqnarray}
The bare tunnel-matrix element $t$ is thus modified by the time dependent vector potential $\bm A(t)$ as 
\begin{equation}
t_{\bm R,\bm R+\bm\delta} = t\exp\left( i2\pi\frac e h \int_{\bm R}
^{\bm R+\delta}d\,\bm l\cdot\bm A(t)\right)=
 t\exp\left( i2\pi\frac e h \bm A(t)\cdot \bm \delta\right)
\,.
\end{equation}

The current operator in the presence of the vector potential is defined as
\begin{equation}
\bm j = -\nabla_{\bm A}H\,.
\end{equation}
In linear response one needs the current operator up to first order in
$\bm A(t)$, which gives
\begin{eqnarray}
\bm j = t\sum_{\bm R,\sigma,\bm \delta}\Big(
\frac {ie}{\hbar}
&(&\bm u_x\delta_x + \bm u_y\delta_y)
a_{\sigma}^{\dagger}
(\mathbf{R})b_{\sigma}(\mathbf{R}+\bm\delta)\nonumber\\
-\frac {e^2}{\hbar^2}&(&\bm u_xA_x\delta_x^2+\bm u_yA_y\delta_y^2)
a_{\sigma}^{\dagger}
(\mathbf{R})b_{\sigma}(\mathbf{R}+\bm\delta)+H.c.
\Big)\,.
\end{eqnarray}
The Maxwell equation 
\begin{equation}
\bm E=-\frac {\partial \bm A(t)}{\partial\,t}
\end{equation}
allows for a simple relation between the vector potential and the
electric field, which is needed for the calculation of the electrical
conductivity.  If one introduces the Fourier representation for the
operators as
\begin{equation}
a_\sigma(\bm R) = \frac 1{\sqrt{N_c}}\sum_{\bm k}e^{-i\bm k\cdot \bm R}a_\sigma(\bm k)\,,
\end{equation}
and the equivalent expression for the operators $b(\bm R)$, the Hamiltonian,
for $\bm A=0$, reads

\begin{eqnarray}
H  =  -t\sum_{\bm k,\sigma }[\phi(\bm k)
a_{\sigma}^{\dagger}(\bm k)b_{\sigma}(\bm k)
+\textrm{H.c.}]
\,,
\label{eq:Hk}
\end{eqnarray}
where
\begin{equation}
\phi(\bm k)=\sum_{\bm \delta}e^{-i\bm k\cdot\bm \delta}\,.
\end{equation}
It proves useful to redefine $\phi(\bm k)$ as 
\begin{equation}
\phi(\bm k)=\sum_{\bm \delta}e^{-i\bm k\cdot(\bm \delta-\bm \delta_3)}\,,
\end{equation}
and introduce the transformation 
$a^\dag_{\bm k}\rightarrow e^{-i\bm k\cdot\bm \delta_3 } a^\dag_{\bm k}$.
After these transformations, the single particle Green's functions for the 
Hamiltonian (\ref{eq:Hk}) are given by

\begin{eqnarray}
G_{AA}^0(\omega_n,\bm k) = \frac {i\omega_n+\mu/\hbar}
{(i\omega_n+\mu/\hbar)^2-t^2\vert\phi(\bm k)\vert^2/\hbar^2 }\,,\\
G_{BA}^0(\omega_n,\bm k) = \frac {-t \phi^\ast (\bm k)/\hbar}
{(i\omega_n+\mu/\hbar)^2-t^2\vert\phi(\bm k)\vert^2/\hbar^2 }\,,\\
G_{BB}^0(\omega_n,\bm k) = \frac {i\omega_n+\mu/\hbar}
{(i\omega_n+\mu/\hbar)^2-t^2\vert\phi(\bm k)\vert^2/\hbar^2 }\,,\\
G_{AB}^0(\omega_n,\bm k) = \frac {-t\phi (\bm k)/\hbar}
{(i\omega_n+\mu/\hbar)^2-t^2\vert\phi(\bm k)\vert^2/\hbar^2 }\,,
\end{eqnarray}
where we also introduced the chemical potential $\mu$.

\section{Current operator and Kubo formula}
Let us now concentrate on the current operator $j_x$, which is composed of the paramagnetic and diamagnetic contribution
$j_x = j_x^P+A_x(t)j_x^D$, each of them given by
\begin{equation}
j^P_x = -\frac {itea}{2\hbar}\sum_{\bm k,\sigma}
[(\phi(\bm k)-3)a_{\sigma}^{\dagger}(\bm k)b_{\sigma}(\bm k) - 
(\phi^\ast(\bm k)-3)b^\dag_{\sigma}(\bm k)
a_{\sigma}(\bm k)
]\,,
\end{equation}
and 
\begin{equation}
j^D_x = -\frac {te^2a^2}{4\hbar^2}\sum_{\bm k,\sigma}
[(\phi(\bm k)+3)a_{\sigma}^{\dagger}(\bm k)b_{\sigma}(\bm k) + 
(\phi^\ast(\bm k)+3)b^\dag_{\sigma}(\bm k)
a_{\sigma}(\bm k)
]\,.
\end{equation}

The Kubo formula for the conductivity is given by\cite{Paul03} 
\begin{equation}
\sigma_{xx}(\omega) = \frac {< j^D_x>}{iA_s(\omega + i0^+)}+
\frac {\Lambda_{xx}(\omega + i0^+)}{i\hbar A_s(\omega + i0^+)}\,,
\end{equation}
with $A_s=N_cA_c$ the area of the sample, and $A_c$ the area of the unit cell,
from which it follows that
\begin{equation}
\Re\sigma(\omega) = D\delta(\omega) + \frac {\Im \Lambda_{xx}(\omega + i0^+)}
{\hbar\omega A_s}\,,
\end{equation}
where $D$ is the charge stiffness which reads
\begin{equation}
D= -\pi \frac {<j^D_x>}{A_s} -\pi\frac {\Re \Lambda_{xx}(\omega + i0^+) }
{\hbar A_s}\,.
\end{equation}
The incoherent contribution to the conductivity $\Lambda_{xx}(\omega + i0^+)$ is obtained from $\Lambda_{xx}(i\omega_n)$,
with this latter quantity defined  as 
\begin{equation}
\Lambda_{xx}(i\omega_n) = \int_0^{\hbar\beta}d\,\tau e^{i\omega_n\tau}
<T_{\tau} j^P_{x}(\tau)j^P_x(0)>\,.
\end{equation}
The $f-$sum rule, giving the oscillator strength, reads
\begin{equation}
\int^\infty_0\sigma_{xx}(\omega)d\,\omega = -\pi \frac {<j^D_x>}{2A_s}\,,
\end{equation}
where we have used $\int_0^\infty\delta(\omega)d\omega = 1/2$, see also Ref. \cite{Gusynin07}.

Within this model, the several quantities read
\begin{equation}
<j^D_x> = -\frac {te^2a^2}{\hbar^2}\sum_{\bm k}\vert\phi(\bm k)\vert[n_F(-t\vert\phi(\bm k)\vert-\mu)
-n_F(t\vert\phi(\bm k)\vert-\mu)]\,,
\end{equation}
\begin{equation}
\Re \Lambda_{xx}(0 + i0^+)=\frac {te^2a^2}{8\hbar} \sum_{\bm k}\frac {f[\phi(\bm k)]}
{\vert\phi(\bm k)\vert} 
[n_F(-t\vert\phi(\bm k)\vert-\mu)
-n_F(t\vert\phi(\bm k)\vert-\mu)]\,,
\end{equation}

\begin{eqnarray}
\Im \Lambda_{xx}(\omega + i0^+)&=&\frac {t^2e^2a^2}{8\hbar^2} \sum_{\bm k}f[\phi(\bm k)] 
[n_F(-t\vert\phi(\bm k)\vert-\mu)
-n_F(t\vert\phi(\bm k)\vert-\mu)]\nonumber\\
&\times&[\pi \delta (\omega -2t\vert\phi(\bm k)\vert/\hbar) -
\pi \delta (\omega +2t\vert\phi(\bm k)\vert/\hbar)
]
\,,
\end{eqnarray}
and
\begin{equation}
f[\phi(\bm k)] = 18-4\vert\phi(\bm k)\vert^2 + 18 \frac {[\Re\phi(\bm k)]^2-[\Im \phi(\bm k)]^2}
{\vert\phi(\bm k)\vert^2}\,.
\end{equation}
The above formulas were derived using the fact that
\begin{eqnarray}
\sum_{\bm k}\phi(\bm k)g(\vert\phi(\bm k)\vert)&=&
\sum_{\bm k}\phi^\ast(\bm k)g(\vert\phi(\bm k)\vert)
\nonumber\\
&=&
\frac 1 3
\sum_{\bm k}\vert\phi(\bm k)\vert^2 g(\vert\phi(\bm k)\vert)
\end{eqnarray}
where $g(\vert\phi(\bm k)\vert)$ is an arbitrary function of the
absolute value of $\phi(\bm k)$, and the fact that in the
Dirac cone approximation one has
\begin{equation}
\frac {\phi^2(\bm k)}{\vert \phi(\bm k)\vert ^2}\simeq
e^{i2\pi/3}[\cos(2\theta)-i\sin(2\theta)]\,,
\end{equation}
and a similar expression for the complex conjugate expression 
$[\phi^\ast(\bm k)]^2/\vert \phi(\bm k)\vert ^2$. One can also show that the following relation holds true 
\begin{equation}
\sum_{\bm k} \vert\phi(\bm k)\vert = \frac 1 8 \sum_{\bm k} \frac {f[\phi(\bm k)]}
{\vert\phi(\bm k)\vert} \,,
\end{equation}
which proves that the charge stiffness is zero at zero
temperature and for zero chemical potential (half-filling). As a
consequence, the system can only show d.c. conductivity at half
filling and zero temperature if
\begin{equation}
\sigma_{\textrm{d.c.}}=
\lim_{\omega\rightarrow 0} \frac {\Im \Lambda_{xx}(\omega + i0^+)}
{\hbar\omega A_s}
\end{equation} 
is finite.

The calculation is simple to do within the Dirac cone approximation.
The term in $f[\phi(\bm k)] $
proportional to $[\Re\phi(\bm k)]^2-[\Im \phi(\bm k)]^2$ gives zero in this approximation.
Let us first introduce the density of states per unit area $\rho(\epsilon)$
\begin{equation}
\rho(\epsilon) = \frac 1 {4\pi^2}\int_0^{q_c}2\pi qdq \delta(\epsilon -\frac 3 2 a q) = 
\frac {4\epsilon}{18\pi a^2}\,. 
\end{equation}
Using $\rho(\epsilon)$ we can calculate
\begin{eqnarray}
\frac 1{A_s}\sum_{\bm k}f[\phi(\bm k)]\delta(\omega-2t|\phi(\bm k)|/\hbar)&\simeq&\int_0^{\epsilon_c}d 
\epsilon
\rho(\epsilon)
(18-4\epsilon^2)\delta(\omega-2t\epsilon/\hbar)\nonumber\\
&=&\frac {\hbar^2}{\pi a^2t^2}\omega \left[
1-\left(
\frac {\hbar\omega}{3\sqrt 2 t}
\right)^2
\right]\,.
\end{eqnarray}
The contribution just computed corresponds to the value of a single
Dirac cone. The optical conductivity is (the two Dirac cone contributions
included)
\begin{equation}
\sigma(\omega) = \frac{\pi}{2}\frac {e^2}{h} \left[
1-\left(
\frac {\hbar\omega}{3\sqrt 2 t}
\right)^2
\right]\,,
\end{equation}
which gives a $d.c.$ value of
\begin{equation}
\sigma_{\textrm{d.c.}}= \frac{\pi}{2}\frac {e^2}{h}.
\label{sigmazero}
\end{equation}
This is the result which one obtains if a finite 
damping term is not included.\cite{Ryu06}

\begin{figure}[t]
\begin{center}
\includegraphics*[angle=0,width=0.8\linewidth]{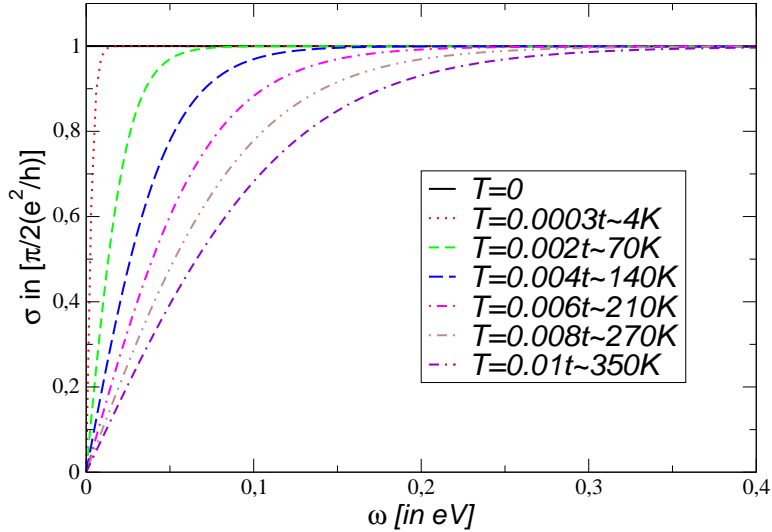}
\caption{The optical conductivity as function of frequency is plotted for various temperatures (for $t=3$eV). \label{fig1}}
\end{center}
\end{figure}

The above result holds only at zero temperature. For finite temperature,
which is the case in any experiment, the conductivity, at half filling and within the Dirac cone approximation, 
is
(note that due to particle-hole symmetry the chemical
potential is zero for any temperature)
\begin{equation}
\sigma(\omega,T) = \frac{\pi}{2}\frac {e^2}{h} \left[
1-\left(
\frac {\hbar\omega}{3\sqrt 2 t}
\right)^2
\right]\tanh \left(
\frac {\hbar\omega}{4k_BT}
\right)\,,
\end{equation}
leading to 
\begin{equation}
\sigma_{\textrm{d.c.}}(T)= 0\,,
\end{equation}
which should be interpreted as the correct result
at zero frequency instead of result (\ref{sigmazero}). This is shown in Fig. \ref{fig1}, 
where the optical conductivity as function of frequency is plotted for 
various temperatures for $t=3$eV. Only the $T=0$ curve yields a finite dc conductivity whereas for finite temperature all curves eventually become zero for $\omega\to0$. E.g. at $T=70K$, the universal value is only reached for frequencies $\omega>0.075$eV, at $T=4K$ for $\omega>0.01$eV. 

\section{Summary}
Starting from the tight-binding model, we have calculated the conductivity of clean gaphene as function of the frequency and temperature. Depending on the order of the limits $\omega\to0$ and $T\to0$, we obtain either the finite value $\sigma=\frac{\pi}{2}\frac{e^2}{h}$ or  zero in agreement with field-theoretical calculations.\cite{Ryu06} The universal value of the conductivity of a clean graphene at finite frequencies was recently confirmed experimentally.\cite{Nair}

Our calculations start from the tight-binding model to complement the current discussion on the conductivity of clean graphene and we believe that the explicit calculation based on the lattice model will be useful to a wide audience. An obvious extension is to include the next-nearest neighbor coupling and corrections to the Dirac cone approximation.\cite{SPG08} Also, the effect of out-of-plane phonons\cite{Sta08} and other scattering mechanisms\cite{Peres08} on the optical conductivity of clean graphene can easily be assessed within the presented formalism.
\section*{Acknowledgements}
N.~M.~R.~Peres thanks the ESF Science Programme INSTANS
 2005-2010, and FCT under the grant PTDC/FIS/64404/2006.
This work has also been supported by MEC (Spain) through Grant
 No. FIS2004-06490-C03-00, by the European Union, through contract
 12881 (NEST), and the Juan de la Cierva Program (MEC, Spain). 


\begin{thebibliography}{99}
\bibitem{FootnoteDeg}
We always include the four channels of graphene in the results of the conductivity, two for the spin-degeneracy and two for the valley-degeneracy.
\bibitem{FootnoteChemPot}
The conductivity of doped (clean) graphene is infinite as in the case of a clean metal.
\end{thebibliography}
\end{document}